BIOPHYSICS

# Positronium imaging with the novel multiphoton PET scanner

Paweł Moskal[1,2]*, Kamil Dulski[1,2]*, Neha Chug[1,2], Catalina Curceanu[3], Eryk Czerwiński[1,2], Meysam Dadgar[1,2], Jan Gajewski[4], Aleksander Gajos[1,2], Grzegorz Grudzień[5], Beatrix C. Hiesmayr[6], Krzysztof Kacprzak[1], Łukasz Kapłon[1,2], Hanieh Karimi[1,2], Konrad Klimaszewski[7], Grzegorz Korcyl[1], Paweł Kowalski[7], Tomasz Kozik[1], Nikodem Krawczyk[1,2], Wojciech Krzemień[8], Ewelina Kubicz[1,2], Piotr Małczak[9], Szymon Niedźwiecki[1,2], Monika Pawlik-Niedźwiecka[1,2], Michał Pędziwiatr[9], Lech Raczyński[7], Juhi Raj[1,2], Antoni Ruciński[4], Sushil Sharma[1,2], Shivani[1,2], Roman Y. Shopa[7], Michał Silarski[1,2], Magdalena Skurzok[1], Ewa Ł. Stępień[1,2], Monika Szczepanek[1,2], Faranak Tayefi[1,2], Wojciech Wiślicki[7]

In vivo assessment of cancer and precise location of altered tissues at initial stages of molecular disorders are important diagnostic challenges. Positronium is copiously formed in the free molecular spaces in the patient's body during positron emission tomography (PET). The positronium properties vary according to the size of inter- and intramolecular voids and the concentration of molecules in them such as, e.g., molecular oxygen, $O_2$; therefore, positronium imaging may provide information about disease progression during the initial stages of molecular alterations. Current PET systems do not allow acquisition of positronium images. This study presents a new method that enables positronium imaging by simultaneous registration of annihilation photons and deexcitation photons from pharmaceuticals labeled with radionuclides. The first positronium imaging of a phantom built from cardiac myxoma and adipose tissue is demonstrated. It is anticipated that positronium imaging will substantially enhance the specificity of PET diagnostics.

## INTRODUCTION

Personalized treatment, which is aimed at providing tailored medical treatment for individual patients, requires diagnostic methods that have a high sensitivity and specificity. Positron emission tomography (PET) meets these requirements to a large extent. PET is one of the most advanced molecular imaging methods, and it enables the detection of tissue alterations on a molecular level before the alterations evolve into functional and, later, morphological disorders (1–3). When performing diagnostics with PET, the patient is administered with a pharmaceutical that is labeled with a radioactive isotope that emits positrons (e.g., $^{18}$F). Positrons thermalize (lose energy) in the tissue and annihilate with electrons, predominantly into two photons. These photons are registered using the PET scanner and are used for the reconstruction of the image for the density distribution of the annihilation points. This distribution is correlated with the metabolism rate of the administered pharmaceutical; this information is used to determine locations of regions with an altered metabolic activity.

Recent developments in total-body PET systems (4–7) have significantly enhanced the PET diagnostic specificity to disentangle cancerous, inflamed, and infected tissues (8, 9). This is achieved by simultaneously enabling a dynamic metabolic rate and kinetic parametric imaging of all the organs, tissues, and cells throughout the human body (10, 11). However, there is still a scope for the quantitative improvement of the PET diagnosis specificity for the disease assessment. This is because the photons that originate from the decay of the radioisotopes and the annihilation of the positron with an electron in the cells can carry more information about the molecular environment than used in the PET diagnostics that are currently being used. The state-of-the-art PET systems (12) [including uEXPLORER, the first total-body PET (13)] use only the information about the location of the positron-electron annihilation and are insensitive to the processes that influence the positron annihilation inside the molecules, which may provide information about the alteration of the tissues at the molecular level (14). In the tissue, positron-electron annihilation may proceed directly ($e^+e^- \rightarrow$ photons) or via the formation of the intermediate positron-electron–bound state, referred to as a positronium ($e^+e^- \rightarrow$ positronium $\rightarrow$ photons). During PET scanning, almost 40% of the positron annihilations occur through the formation of positronium (15, 16), which may be formed in human tissues in the intra- and intermolecular spaces. Hence, positronium has not been used for medical diagnostics.

The positronium is unstable and decays into photons. In 25% of the cases, a short-lived parapositronium is formed, and in 75% of the cases, an orthopositronium (o-Ps) is formed. Parapositronium decays in vacuum predominantly into two photons, and o-Ps decays into three photons, with the mean lifetimes of these states in vacuum equal to 125 ps and 142 ns, respectively (17). In the medium, the mean lifetime of o-Ps is shortened significantly. This is because, in addition to the self-annihilation of o-Ps, additional processes occur, e.g., annihilation of a positron from o-Ps with an electron from the surrounding molecule (pickoff process) or interaction of o-Ps with oxygen or other biomolecules leading to the conversion of o-Ps to parapositronium and its subsequent self-annihilation into two photons. The mean lifetime of o-Ps in a patient's body varies from approximately 1.8 ns in water molecules to approximately 4 ns in skin cells (*18*). Because of the pickoff and conversion processes, the mean lifetime of o-Ps is highly sensitive to the size of inter- and intramolecular voids (free volume between the atoms) and to the concentration of biomolecules in them. As a result, it can provide information about the disease progression in an initial stage (*14*). Several studies have investigated the changes of the positronium properties during dynamic processes while applying model and living biological systems (*19*–*32*). It has been found that there are differences in the positronium mean lifetimes and the production probabilities in healthy and cancerous tissues, indicating that these parameters can be used as indicators for in vivo cancer classification. To exploit these environmentally modified properties of positronium as diagnostic biomarkers for conducting an in vivo tissue pathology assessment, the properties of the positronium atoms need to be determined in a spatially resolved manner (*1*, *14*, *33*). The novel concept of this method, called positronium imaging (*33*, *34*), and feasibility studies based on computer simulations have been reported in our recent publications (*37*, *38*). This article presents the experimental implementation of the method and the first-ever positronium mean lifetime image, which was determined simultaneously with a standard PET image, using a specially designed PET scanner that overcomes the limitations of the state-of-the-art PET system. The scanner, which is referred to as the Jagiellonian PET (J-PET), was constructed and made operational at the Jagiellonian University in Poland (*35*, *36*, *39*–*43*). As an example of the application of the positronium imaging method, this study presents a positronium image that was determined for a phantom comprising cardiac myxoma and adipose tissue from patients (bioethical consent number 1072.6120.123.2017).

[1]Faculty of Physics, Astronomy, and Applied Computer Science, Jagiellonian University, Łojasiewicza 11, 30-348 Kraków, Poland. [2]Total-Body Jagiellonian-PET Laboratory, Jagiellonian University, Kraków, Poland. [3]INFN, Laboratori Nazionali di Frascati, Frascati, Italy. [4]Institute of Nuclear Physics, Polish Academy of Sciences, Kraków, Poland. [5]Department of Cardiovascular Surgery and Transplantology, John Paul II Hospital, Kraków, Poland. [6]Faculty of Physics, University of Vienna, Vienna, Austria. [7]Department of Complex Systems, National Centre for Nuclear Research, Otwock-Świerk, Poland. [8]High Energy Physics Division, National Centre for Nuclear Research, Otwock-Świerk, Poland. [9]2nd Department of General Surgery, Jagiellonian University Medical College, Kraków, Poland.
*Corresponding author. Email: p.moskal@uj.edu.pl (P.Mo.); kamil.dulski@doctoral.uj.edu.pl (K.D.)

## RESULTS
### Positronium detection by the J-PET detector

Figure 1 displays the main results from this study, which include the positronium mean lifetime image and the image analog to the standardized uptake value (SUV). It should be stressed that, in this study, the annihilation rate is determined by the activity of the applied $^{22}$Na radionuclide and not by the pharmaceutical uptake rate as it is in the in vivo PET imaging. Both of these images were obtained simultaneously for the phantom, which comprised tissues that were operated from two patients (indicated as 1 and 2 in the J-PET tomography chamber). The imaged phantom consisted of four samples inserted into the J-PET tomograph. Two samples consisted of cardiac myxoma tissue (indicated as blue circles), and the other two consisted of adipose tissue (indicated as yellow circles).

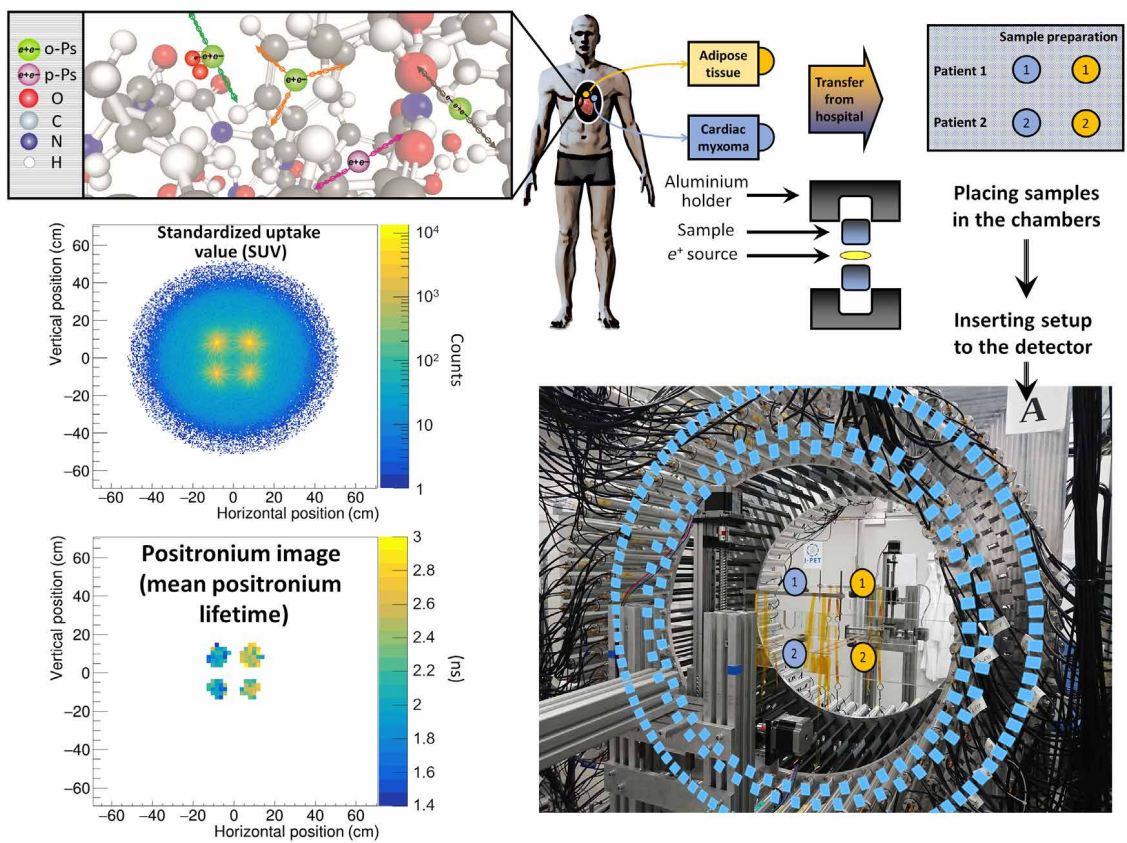

**Fig. 1. The measurement scheme and the determined SUV and positronium images.** The top left panel illustrates a part of a hemoglobin molecule with superimposed schemes, indicating decays of o-Ps (green circles) and parapositronium (p-Ps) (violet circle). o-Ps may undergo self-annihilation (pink arrows), pickoff process (gray arrows), or conversion to parapositronium, e.g., by interacting with oxygen molecule (green arrows). The top right panel shows the first part of the experimental workflow. Four samples from two patients were collected and divided into two classes: cardiac myxoma and adipose tissue. Each sample was inserted into a holder with the radioactive $^{22}$Na source, which was inserted into the J-PET detector. The blue and yellow circles show the locations of the samples during the measurement. Methods to reconstruct the image analogs to that of the SUV image (annihilation rate distribution) and positronium lifetime image are described in Methods. Reconstructed mean o-Ps lifetime in cardiac myxoma (1.9 ns) differs from the mean o-Ps lifetime in adipose tissue (2.6 ns). More comprehensive studies confirming the differences in the o-Ps mean lifetime in the healthy adipose tissue and cardiac myxoma tumor are described in the article to be submitted elsewhere (*44*). Photo credit: Kamil Dulski, Jagiellonian University.

$^{22}$Na isotope was used as a source of positrons. In addition to emitting the positrons, this isotope emits a prompt gamma ray with an energy of 1.27 MeV via the following reaction chain (Fig. 2): $^{22}$Na → $^{22}$Ne* + $e^+$ + $\nu$ → $^{22}$Ne + $\gamma$(1.27 MeV) + $e^+$ + $\nu$. The source was surrounded by tissues and closed in the aluminum holder that was placed in the tomography chamber. The positrons that were emitted from the source penetrated the tissue to a depth of about 1 mm (*45*) and annihilated with the electrons from the molecules that constitute the cells (Fig. 2). In about 40% of the cases, annihilation proceeds by formation of the positronium atoms, which are trapped in the intramolecular spaces (Fig. 1) (*14*). Seventy-five percent of the positronium atoms are produced as long-lived o-Ps. o-Ps has a lifetime that is very sensitive to the size of the voids and the concentration in them of molecules such as oxygen molecules (*19*, *20*). In the intramolecular voids of the tissue (as illustrated in the top left panel of Fig. 1), the o-Ps annihilates about 70 times more often via the pickoff and conversion processes and then via self-annihilation. Therefore, as a result of the o-Ps annihilation in the tissue, the emission of two photons is about 70 times more likely than the emission of three photons (*37*, *38*, *46–48*). Therefore, for the first demonstration of the positronium imaging, this study selected triple coincidence events corresponding to the registration of two annihilation photons and one prompt gamma. The time and position of the interaction of the annihilation photons in the plastic strips of the J-PET tomograph were used to reconstruct the position and time of the annihilation (Figs. 2 and 3). Meanwhile, the time and position of the interaction of the prompt gamma were used to reconstruct the emission time of the prompt gamma. Because of the short time of the positron thermalization [in the order of 10 ps (*49*)] and a short time of $^{22}$Ne deexcitation [on average, it is about 3 ps (*50*)], the time of the prompt gamma emission is within tens of picoseconds equivalent to the time of the positron emission and the time of the positronium formation.

### Positronium imaging
The reconstructed density distribution of the position of the annihilation points (Fig. 1) constitutes the image of the annihilation rate [an analog of the standard uptake value (SUV image)]. In the studied case, it reflects the geometrical configuration of the tissues in the tomographic chamber (four samples are set at the vertices of the square) and the activity of the applied $^{22}$Na sources (for details, please see Table 1). For each voxel of the SUV image, a distribution of ($\Delta t$) the difference between the annihilation time and the time of the positron emission was determined. Each $\Delta t$ distribution was fitted using the dedicated PALS Avalanche program (*51*, *52*) to determine the mean lifetime for the o-Ps on a voxel-by-voxel basis. An image for the mean lifetime of o-Ps is shown in Fig. 1. It indicates that the mean lifetime of the o-Ps in cardiac myxoma is approximately 1.9 ns, and it differs significantly from the mean lifetime of the o-Ps in adipose tissues, which is approximately 2.6 ns.

The SUV and positronium images presented in Fig. 1 were reconstructed on the basis of the data that were collected with the J-PET tomograph (this is described in detail in Methods). The J-PET tomograph consists of 192 plastic scintillator strips (covered in a light-tight black foil that is visible in the photograph), and it is arranged into three concentric layers, as it is visualized by superimposed rings from the blue rectangles. Each scintillator is read out by the photomultipliers that are optically connected at the ends (*39*). The signals from the photomultipliers are sampled in the voltage domain by the newly developed programmable electronics (*41*). The data were collected with the continuous readout mode (*43*), and the filtration of events was performed off-line with a specialized software program (*53*).

### DISCUSSION
The mean o-Ps lifetime image shown in Fig. 1 corresponds to the first-ever positronium image. It was determined using the J-PET tomograph. By overcoming the limitations of the current PET systems, it demonstrates the possibility of simultaneously achieving PET SUV imaging and positronium imaging using the PET detectors. The presented results also demonstrate the potential of positronium imaging in enhancing the specificity in the PET diagnostics. Visible and significant differences between the o-Ps lifetime in cancer and healthy tissues are observed, as shown in the example of the cardiac myxoma and adipose tissues. The observed difference in the mean lifetime of o-Ps in cardiac myxoma and adipose tissue is equal to about 700 ps, while the differences due to the changes of the concentration of molecules are expected at the level of less than 10 ps as demonstrated for the case of oxygen molecules in the recent article [equation 9 in (*20*)]. Thus, we may assume that the observed difference will also remain significant under the in vivo conditions, although the concentration of biomolecules differs between dead and alive cells. The comprehensive studies of these phenomena will have to be carried out to answer of how to interpret changes in the positronium image in individual organs or tissue types.

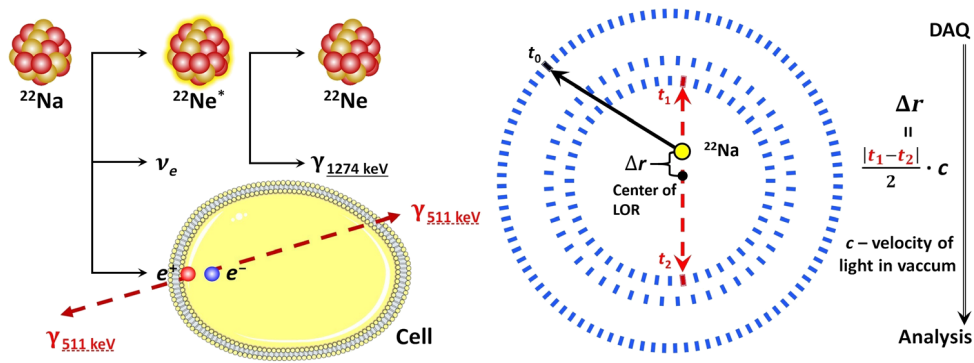

**Fig. 2. Experiment workflow.** The $^{22}$Na radionuclide emitting positron and prompt gamma is used as a source of positrons. They can annihilate with the electrons from the molecules of the cell. The annihilation photons and prompt gamma interact in the scintillator strips (blue rectangles) of the J-PET detector; this results in the production of signals that are sampled by the front-end electronics (*41*), digitized, and collected by a data acquisition (DAQ) system (*43*). A single event of interest comprises the information of the position and time ($x_i, y_i, z_i, t_i$) for each registered photon. The events, in which the detector registers two annihilation photons and one prompt gamma, are selected for further analysis. The reconstructed positions of the annihilation photons, ($x_1,y_1,z_1$) and ($x_2,y_2,z_2$), enable the reconstruction of the so-called line of response (LOR), indicated by the red dashed line, which comprises the point of annihilation (yellow circle). Analysis part of the experiment is given in Fig. 3.

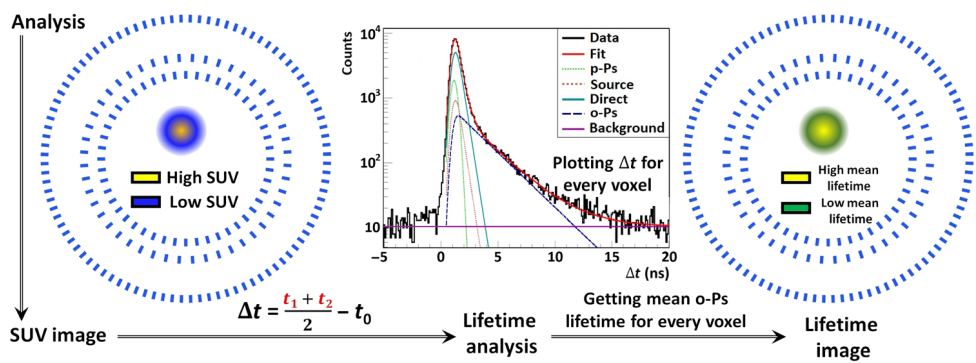

**Fig. 3. Analysis of the experiment.** Experiment conducted as in Fig. 2 was analyzed by the following procedure. The distribution of the reconstructed positions of the annihilations constitutes the SUV image [a pictorial example of this image is shown in (**left**)]. The photons resulting from the deexcitation of the $^{22}$Ne* can be used to calculate the positron annihilation lifetime ($\Delta t$) distribution for every voxel in the SUV image. The decomposition of the $\Delta t$ distribution [an example distribution for cardiac myxoma tissue is shown in (**middle**)] allows us to distinguish components coming from the parapositronium (light green dashed line) and o-Ps (dark blue dashed line). It also distinguishes processes such as direct annihilation of the positron and the electron without producing positronium (direct; turquoise line) and from the annihilations in the source material (source; red dashed line). In addition, the signal from accidental coincidences is visible (background; violet line). The positronium image [shown pictorially in (**right**)] has each voxel filled with the value of the mean o-Ps lifetime, and it is obtained from the decomposition of the $\Delta t$ distribution.

**Table 1. Geometry and activity of the phantom.** The positions of the imaged samples and the activity of the applied $^{22}$Na sources are listed in the second and third columns, respectively. The fourth column indicates number of events in a given spectrum fulfilling all the selection criteria described in the text. In total, out of about $10^7$ collected three-hit events, only 232,357 events were used in the lifetime spectra shown in Fig. 5. The activity of the source, which is normalized to the total number of counts, is given in the last column. It shows that the reconstructed intensity in the SUV image, shown in Fig. 1, is in good agreement with the expectations that are based on the activity of the used $^{22}$Na sources.

| Sample | Position X, Y, and Z (cm) | Source activity (MBq) | Total number of counts | Intensity normalized to activity |
|---|---|---|---|---|
| Cardiac myxoma 1 | −8.1, 8.1, 0 | 0.345 | 65 636 | 0.988 |
| Cardiac myxoma 2 | −8.1, −8.1, 0 | 0.238 | 45 493 | 0.993 |
| Adipose tissue 1 | 8.1, 8.1, 0 | 0.393 | 76 651 | 1.013 |
| Adipose tissue 2 | 8.1, −8.1, 0 | 0.230 | 44 577 | 1.007 |

The time-of-flight resolution of the constructed J-PET prototype amounts to 460 ps [full width at half maximum (FWHM) for the coincidence resolving time (CRT)] and corresponds to the spatial resolution of 6.9 cm (FWHM) in the image. Therefore, for the first images that were presented in this article, we used a rather large voxel size of 2 cm by 2 cm by 2 cm. However, the continuous advancement of the time resolution for the PET scanners (*54–57*) has promising perspectives to improve this resolution in the future. The best current clinical PET systems can achieve a time resolution of 210 ps (*58*), and the laboratory systems reach the CRT in 30 ps (*59*). This could result in a further improvement, even up to 10 ps (*54*, *55*, *57*); thus, it could create positronium images with a spatial resolution of 0.15 cm. The sensitivity for the positronium imaging of the constructed J-PET demonstrator is only 0.0016%. However, the advent of the total body PET systems and, in particular, the J-PET total-body PET system (*1*) can enhance the sensitivity for total body positronium imaging, such that it will be more than double in comparison to 2γ PET imaging, corresponding to the current PET scanners that have a 20-cm axial field of view (*1*). For the CRT of about 500 ps, as presently achieved by the uEXPLORER total-body PET, the point spread function of the positronium image would equal to 30 mm (radial) and 7 mm (axial) (*38*).

Last, it is important to stress that the information that is comprised in the positronium images concerns the size of inter- and intramolecular voids and the concentration in them of molecules, e.g., oxygen molecules. Therefore, it is qualitatively different from the anatomical and morphological images that are obtained by computed tomography and magnetic resonance imaging. It is also qualitatively different from the metabolic rate images that are obtained with the current PET systems. Therefore, when using positron imaging in clinical practice, the specificity of PET diagnostics in disease assessment will increase. However, establishing the quantitative correlations between the positronium properties in the tissue and the type and degree of tissue pathology still requires further research (*1*, *14*).

The method can be also applied to the regular PET system based on crystal scintillators. Note that crystal-based PET systems with an axial field of view of 60 cm may achieve sensitivity for positronium imaging comparable to the standard 2γ PET imaging with 20-cm-long PET scanners (*1*). New commercial tomographs offered [and already installed in clinics, e.g., by Siemens (*60*) or United Imaging companies (*61*)] had an axial field of view longer than 100 cm.

## METHODS
### Source of positrons
A $^{22}$Na radionuclide was used as a positron source. $^{22}$Na undergoes $\beta^+$ decay by emitting an electron neutrino ($\nu_e$) and a positron ($e^+$), which can annihilate with an electron ($e^-$) that comes from the cell (Fig. 2, yellow), thus resulting in the emission of two photons with an energy of 511 keV that are emitted in the opposite directions. After the $\beta^+$ emission, $^{22}$Na is transformed to the excited state of the $^{22}$Ne* isotope, which then deexcites with the emission of a prompt gamma ray with an energy of 1275 keV. The emitted photons can be registered by the J-PET detector when they deposit the energy in scintillators (blue rectangles in Fig. 2) through the Compton effect.

### Preparation of the samples and the setup
The samples of the cardiac myxoma and adipose tissue were extracted from two patients from the John Paul II Hospital in Kraków (bioethical consent number 1072.6120.123.2017). After transportation to the Faculty of Physics, Astronomy, and Applied Computer Science of the Jagiellonian University, the samples were fixed in 10% formalin. Every fixed sample was cut into two pieces to cover the $^{22}$Na source from both sides, as shown in Fig. 1. The samples with radioactive sources were inserted into plastic chambers and placed on a Kapton scaffolding in the J-PET detector, equidistant from the center, as shown in Fig. 1. The activities of the applied $^{22}$Na sources are listed in Table 1.

### J-PET detector
The J-PET detector photograph is shown in Fig. 1, and transversal cross section is shown schematically in Fig. 2. It is constructed using 192 plastic scintillator strips that are arranged in three cylindrical and concentric layers. These include the first layer, which has 48 scintillators and is 42.5 cm away from the center; the second layer, which has 48 scintillators and is 46.75 cm away from the center; and the third layer, which has 96 scintillators and is 57.5 cm away from the center. EJ-230 plastic scintillators with dimensions of 7 mm by

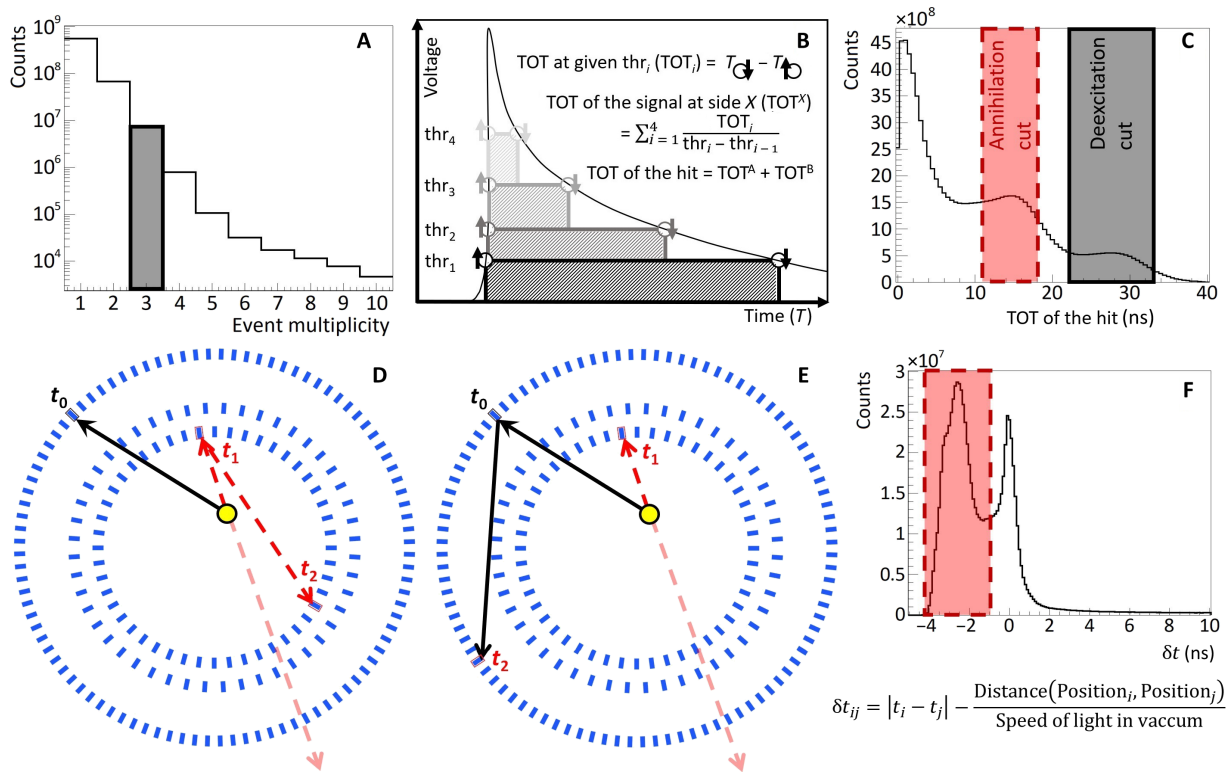

**Fig. 4. Event selection procedure.** The top left panel (**A**) represents the distribution of the multiplicity of the hits. In this analysis, only the events with three hits (signals registered in three scintillator strips) were taken into account. The top middle panel (**B**) describes the signal sampling method. The dots represent the crossing of signals with the preset voltage threshold. Each dot corresponds to the measurement of the time (time stamp) that is digitized and collected by the electronic readouts and DAQ system. The area of the signal is approximated by the weighted mean of the TOT values. In the analysis, as the TOT characterizes the energy loss (*62*), the sum of the TOT values measured at both sides (side A and B) of the scintillator strip is used. The top right panel (**C**) represents the TOT distribution. The Compton edges for 511- and 1275-keV photons are visible at 18 and 33 ns, respectively. The red and gray regions indicate the TOT range used for the identification of the annihilation and prompt photons, respectively. (**D** and **E**) The example of possible background events where one of the annihilation photons was not registered; instead, another annihilation photon (D) (dashed red arrow) or a prompt photon (E) (solid black arrow) was scattered twice. The bottom left panel (**F**) shows the distribution of $\delta t_{ij}$ for a given pair of hits ($i,j$). The events corresponding to the scatterings in the detector are suppressed by the selection of the $\delta t$ values in the range indicated in red.

19 mm by 500 mm are used as an active part of the detector (Eljen Technology, https://eljentechnology.com/). Both ends of the long axis of the scintillator were connected to the vacuum photomultipliers, which form a single detection module (*35, 39, 42*). The light that is generated in the scintillators is converted to electric signals using the Hamamatsu R9800 vacuum photomultipliers (Hamamatsu, https://www.hamamatsu.com/us/en/index.html). The electric signals from the photomultipliers are probed at four thresholds using dedicated front-end electronics (*42*) and converted to the binary data by the novel trigger-less data acquisition (DAQ) system (*43*).

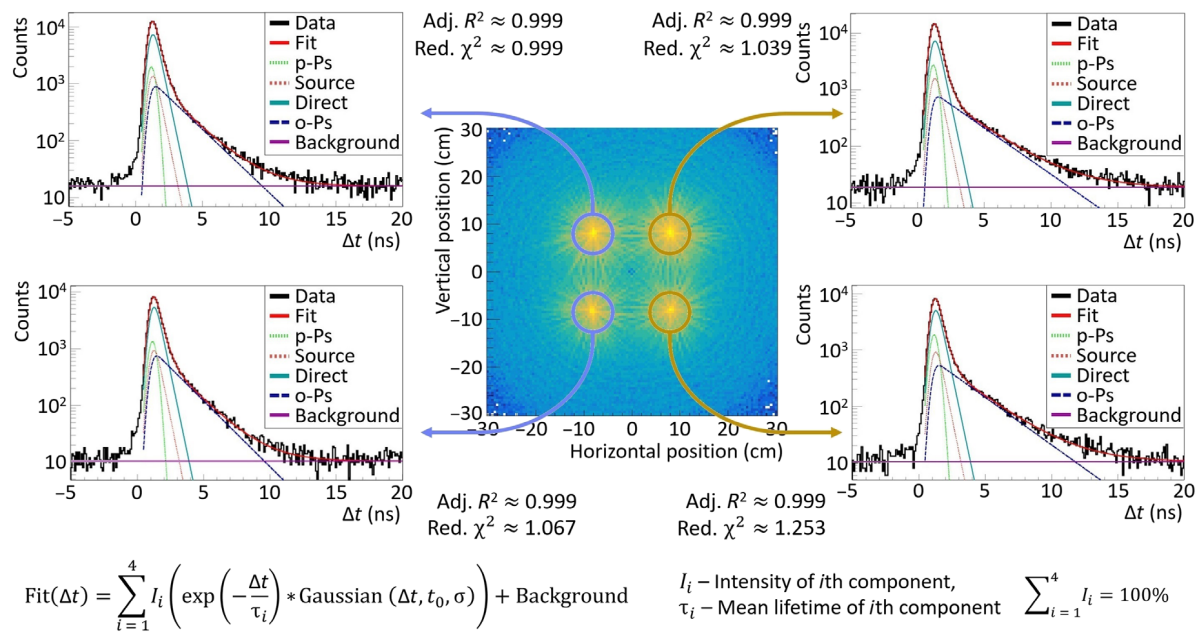

**Fig. 5. Positron annihilation lifetime spectra.** The middle panel shows the SUV image (as in Fig. 1) in the range from −30 to 30 cm. On the left and right sides of the SUV image, the exemplary positron lifetime spectra ($\Delta t$) are shown for the regions that are marked as blue and yellow circles. The positron annihilation components were estimated by fitting these spectra with the PALS Avalanche software program (*51, 52*), which is dedicated for the J-PET detector. The exponential tail is similar for the samples on the right (o-Ps crossing background line near 12 ns) and on the left (o-Ps crossing the background line near 10 ns) separately. The model function is given by Fit($\Delta t$), which is a convolution of the exponential and Gaussian functions. Fits are in a very good agreement with the data confirmed by the values of the resulting adjusted $R^2 = 0.999$ for all of the fits and reduced $\chi^2$ values: (top left) 0.999, (bottom left) 1.067, (top right) 1.039, and (bottom right) 1.253. The resulting fitted parameters of each component for each sample are shown in Table 2. The amplitudes of the individual components for the same type of the samples (myxoma or adipose) differ by only 7% relative value. The mean o-Ps lifetimes for the cardiac myxoma samples are equal to (top left) 1.950 (19) ns and (bottom left) 1.874 (20) ns with an intensity of (top left) 21.39 (47)% and (bottom left) 23.27 (45)%. For the adipose tissue, the mean o-Ps lifetimes are equal to (top right) 2.645 (27) ns and (bottom right) 2.581 (30) ns with an intensity of (top right) 21.49 (41)% and (bottom right) 21.56 (54)%.

**Table 2. Results from fitting PAL spectra for each sample position shown in Fig. 5.** Parapositronium and direct annihilation mean lifetimes were fixed to values of 0.125 and 0.388 ns, respectively. This resulted in more stable results for fitting the mean lifetime of o-Ps, the only parameter of interest. o-Ps mean lifetime was higher for adipose tissue for both samples. Parapositronium and direct annihilation intensities were in agreement for the same type of the sample, cardiac myxoma or adipose tissue. Values of the o-Ps mean lifetime was very close for a given type of the sample. Fitted models were in good agreement with the data indicated by adjusted $R^2$ and reduced $\chi^2$ values close to 1.

| Parameter name | Cardiac myxoma 1 | Cardiac myxoma 2 | Adipose tissue 1 | Adipose tissue 2 |
|---|---|---|---|---|
| Parapositronium mean lifetime (ns) | 0.125 (fixed) | | | |
| Parapositronium intensity (%) | 13.26 (18) | 12.21 (21) | 17.18 (16) | 17.14 (20) |
| Direct annihilation mean lifetime (ns) | 0.388 (fixed) | | | |
| Direct annihilation intensity (%) | 65.35 (22) | 64.52 (27) | 61.34 (20) | 61.31 (23) |
| o-Ps mean lifetime (ns) | 1.950 (19) | 1.874 (20) | 2.645 (27) | 2.581 (30) |
| o-Ps intensity (%) | 21.39 (47) | 23.27 (45) | 21.49 (41) | 21.56 (54) |
| Adjusted $R^2$ | 0.999 | 0.999 | 0.999 | 0.999 |
| Reduced $\chi^2$ | 0.999 | 1.067 | 1.039 | 1.253 |

## Framework and analysis procedure

The data collected using the J-PET detector were analyzed using a specialized software program called the J-PET Framework (*53*), which is based on the C++ architecture. The analysis was performed in several steps: reconstructing the signals, reconstructing the hits, creating events, and the final event-by-event analysis. The DAQ system collects the times when the electric signal produced in the photomultiplier crosses the system thresholds. The signal is reconstructed when it passes one or more thresholds in a given time window (23 ns) for a given photomultiplier. A hit is generated from a pair of signals that come from two different photomultipliers connected to the same scintillator in a time window of 5 ns. Last, an event is a set of hits that are registered within a time window of 200 ns. A distribution of the hit multiplicity during an event is shown in Fig. 4A.

When reconstructing the positronium image, the signals from two annihilation photons and a prompt photon are required. Therefore, in this analysis, only events with three registered hits were taken into account. The origin of the hits was chosen on the basis of the time-over-threshold (TOT) value, which is an estimate of the deposited energy in a scintillator (*62*). The TOT value was calculated as the weighted mean of the widths of the signals for four different thresholds, in which the weights were selected on the basis of the difference between two consecutive thresholds as shown in Fig. 4B. The TOT distribution can be seen in Fig. 4C. The red area indicates the energy deposition that is close to the Compton edge that corresponds to the 511-keV photon that comes from the positron-electron annihilation, and the black area represents 1275-keV photons from the deexcitation of the source. For the selected events that contain two annihilation hits, the geometry condition is checked (the angle between two annihilation hits), so that the annihilation potentially occurs from the interior of the J-PET detector. All the potential scatterings between the neighboring scintillators were suppressed by the additional geometrical condition in which the angle between the prompt gamma and the annihilation photon hit positions is larger than 15°. In general, the events are filtered to suppress the scatterings and other sources of the background by applying the interaction time, energy deposition, and geometrical relations between the positions of the interactions. The main source of background is formed because of the misidentification of the signals from the photons scattered in the detector as signals from the annihilation photons. An example of the background events is presented in Fig. 4 (D and E). This background is suppressed by testing whether the difference between the times of the two hits is equal to the time required by light to travel the distance between these hit positions. The distribution of this defined test parameter δ$t$ is shown in Fig. 4F. If the hits correspond to the scattered photon, then the value of δ$t$ is close to zero. Therefore, the events with δ$t$ > −1 ns are rejected.

From the selected events, the annihilation position is reconstructed, and the SUV image is created. For each voxel in the SUV image, the positron annihilation lifetime distribution is determined, in which the mean o-Ps lifetime is extracted by the PALS Avalanche software program (*51*, *52*). As an example, four positron annihilation lifetime spectra from four areas that correspond to the four different samples are shown in Fig. 5.

## PALS Avalanche

A dedicated software program for the positron lifetime analysis (*51*, *52*) that is compatible with the J-PET Framework software program allows the positron lifetime spectrum to be decomposed into different components. This software was written in C++, and it is based on the ROOT libraries (*63*). The PALS Avalanche software program enables the fitting of the multiexponential distribution, which is smeared with a linear combination of the Gaussians [Fit(Δ$t$) formula given in Fig. 4]. The decomposition leads to the extraction of the mean o-Ps lifetime in a given voxel. The influence of the Kapton foil that covers each source was estimated as the 10% intense component with a mean lifetime of 0.374 ns. The parapositronium mean lifetime was fixed to a value of 0.125 ns. Since the only parameter of interest was the mean lifetime of o-Ps, it can yield more stable results of the fitting. The resolution function was estimated as a single Gaussian with a σ value of 0.266 ns, which corresponds to a time resolution for a single measured time of 0.154 ns. The resulting parameters of each component are shown in Table 2.

**Acknowledgments:** We would like to acknowledge the technical and administrative support from A. Heczko, M. Kajetanowicz, and W. Migdał. We are thankful to B. Jasińska and M. Gorgol for support in preparing the radioactive sources. **Funding:** This work was supported by the Polish National Center for Research and Development through grant INNOTECH-K1/IN1/64/159174/NCBR/12, the Foundation for Polish Science through the MPD and TEAM/2017-4/39 programs, the National Science Centre of Poland through grant nos. 2017/25/N/NZ1/00861 and 2019/35/B/ST2/03562, the Ministry for Science and Higher Education through grants nos. 6673/IA/SP/2016 and 7150/E-338/SPUB/2017/1, the Jagiellonian University via project CRP/0641.221.2020, the Austrian Science Fund FWF-P26783, and the SciMat Priority Research Area budget under the program Excellence Initiative-Research University at Jagiellonian University. **Author contributions:** Conception and design of the investigations, supervision, funding acquisition, interpretation, and guiding the data analysis: P.Mo. Writing—review and editing: P.Mo. and K.D. Concept and elaboration of the PALS Avalanche software program and data analysis: K.D. Concept and guiding the medical research and supervision of the medical protocols: E.Ł.S. Recruitment and operation of the patients: G.G. Sample preparations: E.K., K.D., E.Ł.S., H.K., and M.D. Elaboration of the Framework software and image reconstruction algorithms: W.K., A.G., K.Ka., S.S., K.D., N.K., K.Kl., P.K., L.R., R.Y.S., and W.W. Programming and commissioning of the DAQ system: G.K. Elaboration of the calibration methods and calibration of the J-PET detector: S.N., K.D., M.Si., and M.Sk. Data collection: E.C., E.K., N.C., and F.T. Building, commissioning, testing all 192 detection units, constructing, taking into operation, and determining the performance characteristics of the full detector system: P.Mo., K.D., N.C., C.C., E.C., M.D., A.G., G.G., J.G., B.C.H., K.Ka., Ł.K., H.K., K.Kl., G.K., P.K., T.K., N.K., W.K., E.K., P.Ma., S.N., M.P.-N., M.P., A.R., L.R., J.R., S.S., S., R.Y.S., M.Si., M.Sk., E.Ł.S., M.Sz., F.T., and W.W. All authors took part in the discussions and interpretation of the data, and they read, corrected, and approved the final manuscript. **Competing interests:** P.Mo. is an inventor on a patent related to this work [patent nos.: (Poland) PL 227658, (Europe) EP 3039453, and (United States) US 9,851,456], filed (Poland) 30 August 2013, (Europe) 29 August 2014, and (United States) 29 August 2014; published (Poland) 23 January 2018, (Europe) 29 April 2020, and (United States) 26 December 2017. The authors declare that they have no other competing interests. **Data and materials availability:** All data needed to evaluate the conclusions in the paper are present in the paper.

Submitted 15 March 2021
Accepted 23 August 2021
Published 13 October 2021
10.1126/sciadv.abh4394

**Citation:** P. Moskal, K. Dulski, N. Chug, C. Curceanu, E. Czerwiński, M. Dadgar, J. Gajewski, A. Gajos, G. Grudzień, B. C. Hiesmayr, K. Kacprzak, Ł. Kapłon, H. Karimi, K. Klimaszewski, G. Korcyl, P. Kowalski, T. Kozik, N. Krawczyk, W. Krzemień, E. Kubicz, P. Małczak, S. Niedźwiecki, M. Pawlik-Niedźwiecka, M. Pędziwiatr, L. Raczyński, J. Raj, A. Ruciński, S. Sharma, Shivani, R. Y. Shopa, M. Silarski, M. Skurzok, E. Ł. Stępień, M. Szczepanek, F. Tayefi, W. Wiślicki, Positronium imaging with the novel multiphoton PET scanner. *Sci. Adv.* **7**, eabh4394 (2021).